\documentclass[sigconf]{acmart}

\usepackage{amssymb}
\usepackage{amsmath}
\usepackage{mathtools}
\usepackage{bm}
\usepackage{multirow}
\usepackage{longtable}
\usepackage{adjustbox}
\usepackage{enumitem}

\usepackage{graphicx}
\usepackage{subcaption}
\usepackage{float}
\usepackage{algorithm}
\usepackage{algpseudocode}
\usepackage{xcolor}
\usepackage[normalem]{ulem}
\usepackage[table]{xcolor}
\useunder{\uline}{\ul}{}
\usepackage{pifont}
\usepackage{ragged2e}
\usepackage{verbatim}
\usepackage{microtype}
\usepackage{setspace}
\usepackage{balance}
\usepackage{url}
\usepackage{hyperref}

\AtBeginDocument{%
  }

\setcopyright{acmlicensed}
\copyrightyear{2018}
\acmYear{2018}
\acmDOI{XXXXXXX.XXXXXXX}
\acmConference[Conference acronym 'XX]{Make sure to enter the correct
  conference title from your rights confirmation email}{June 03--05,
  2018}{Woodstock, NY}
\acmISBN{978-1-4503-XXXX-X/2018/06}

\begin{document}

\title[SPARC]{SPARC: Sequence-aware Progressive Attribute Routing and Compression Framework for Generative Recommendation}

\author{Chang Liu}
\email{shimao.lc@alibaba-inc.com}
\affiliation{%
  \institution{Alibaba Group}
  \city{Beijing}
  \country{China}
}
\authornote{Work done during internship at Alibaba.}

\author{Changfa Wu}
\email{wuchangfa.wcf@alibaba-inc.com}
\affiliation{%
  \institution{Alibaba Group}
  \city{Beijing}
  \country{China}
}
\authornote{Corresponding author.}
\authornote{Equal contribution.} 

\author{Hui Qian}
\email{yihui.qh@alibaba-inc.com}
\affiliation{%
  \institution{Alibaba Group}
  \city{Beijing}
  \country{China}
}

\author{Binbin Cao}
\email{simon.cbb@alibaba-inc.com}
\affiliation{%
  \institution{Alibaba Group}
  \city{Beijing}
  \country{China}
}

\author{Jian Wu}
\email{joshuawu.wujian@alibaba-inc.com}
\affiliation{%
  \institution{Alibaba Group}
  \city{Beijing}
  \country{China}
}

\author{Yuliang Yan}
\email{yuliang.yyl@alibaba-inc.com}
\affiliation{%
  \institution{Alibaba Group}
  \city{Beijing}
  \country{China}
}

\author{Han Zhu}
\email{zhuhan.zh@alibaba-inc.com}
\affiliation{%
  \institution{Alibaba Group}
  \city{Beijing}
  \country{China}
}

\author{Bo Zheng}
\email{bozheng@alibaba-inc.com}
\affiliation{%
  \institution{Alibaba Group}
  \city{Beijing}
  \country{China}
}

\renewcommand{\shortauthors}{Chang Liu et al.}

\begin{abstract}
    Generative recommendation tokenizes items as discrete Semantic IDs (SIDs) and autoregressively generates target items from users' historical SID sequences. Although existing SIDs incorporate multimodal and structured information, they are typically statically assigned and independent of the current interaction context. In industrial scenarios, each behavior also contains heterogeneous attributes, such as category, brand, price, behavior type, and timestamp. Fully expanding these features greatly increases the input length, while directly compressing them into a single representation may prematurely discard context-relevant information.

We propose \textbf{SPARC}, \uline{\textbf{S}}equence-aware \uline{\textbf{P}}rogressive \uline{\textbf{A}}ttribute \uline{\textbf{R}}outing and \uline{\textbf{C}}ompression Framework for Generative recommendation. SPARC first models the sequential dependencies of each field type to obtain context-aware field representations. It then routes the original, contextual, and identity representations of different fields into multiple slots to preserve complementary information under a fixed capacity. Finally, lightweight cross-item interaction integrates the intermediate tokens and compresses each historical item into a single token. Following the principle of contextualizing before compression, SPARC enriches user-history representations without increasing the input length of the generative backbone.

Experiments on industrial Taobao and public Amazon datasets demonstrate that SPARC outperforms strong conventional and generative baselines. Further comparisons with static compression variants show that the improvement of SPARC comes from context-conditioned information retention rather than merely increasing the expressiveness of the compression module.
\end{abstract}



\received{20 February 2007}
\received[revised]{12 March 2009}
\received[accepted]{5 June 2009}

\maketitle

\section{Introduction}
\begin{figure}[]
\centering
\includegraphics[width=\linewidth]{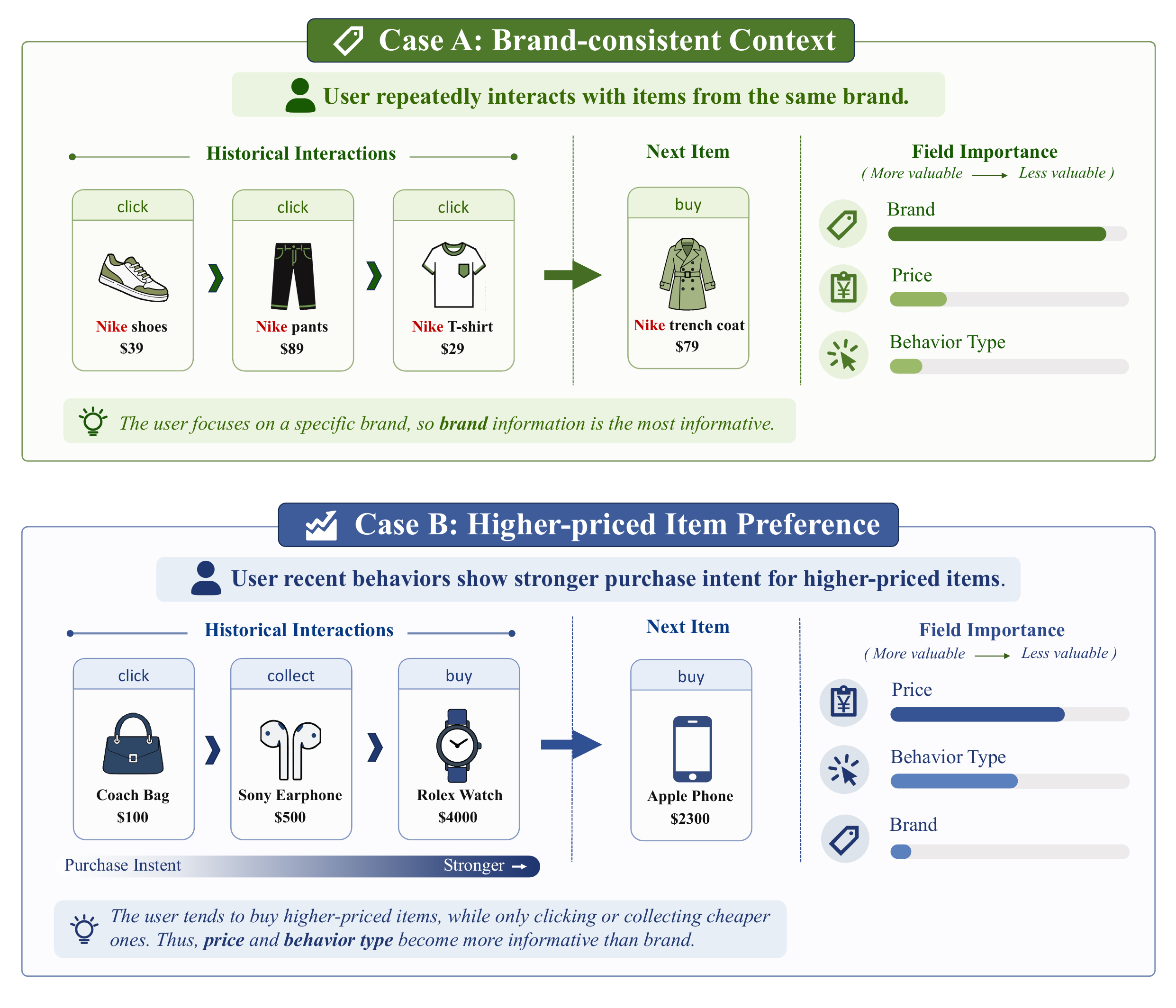}
\caption{Illustration of context-dependent field importance in generative recommendation. Repeated interactions with the same brand make brand information highly predictive, whereas shifts in price range or purchase intent increase the importance of price, behavior type, and interaction time.}
\label{fig:intro_1}
\end{figure}

Generative recommendation has recently attracted increasing attention from both academia and industry~\cite{TIGER,EAGER,ETEGRec,DAS}. Unlike conventional discriminative recommenders that score items from a predefined candidate set, generative recommendation formulates recommendation as the autoregressive generation of item tokens. A typical framework first tokenizes candidate items into discrete Semantic IDs (SIDs), converts a user's interaction history into the corresponding SID sequence, and employs a Transformer-based generative model to predict the SID of the target item~\cite{TIGER,LETTER,Cost,ETEGRec}. By unifying item representation, user modeling, and candidate retrieval within a sequence-generation framework, generative recommendation enables end-to-end prediction over large item spaces and exhibits promising long-tail generalization and industrial applicability~\cite{TIGER,singh2024better,SETRec,DAS}.

Within this paradigm, the quality of SIDs directly affects subsequent user modeling and target-item generation~\cite{Cost,LETTER,ETEGRec,GRID}. Early methods derived continuous semantic representations from item titles and descriptions and discretized them through clustering, vector quantization, or residual quantization~\cite{VQRec,P5,TIGER,Cost}. Subsequent studies incorporated categories, brands, structured attributes, and visual content through multimodal alignment, attribute reconstruction, and auxiliary prediction objectives~\cite{Grace,MMRQVAE,UTGRec,MMQ}. More recent approaches further integrate collaborative relations, behavioral sequences, and temporal signals into SID learning, while some methods explicitly insert behavior- or time-specific tokens into the input sequence~\cite{LETTER,EAGER,DAS,TAFAME,MBSR,Grace}.

Despite these advances, most existing methods still assign each item a globally shared SID~\cite{TIGER,LETTER,Cost,TAFAME}. Regardless of how much textual, visual, structured, collaborative, or sequential information is incorporated during tokenizer training, the same item typically retains an identical token across users and interaction scenarios. Such representations effectively capture stable item semantics and collaborative patterns but cannot fully characterize a specific historical interaction. In real-world industrial systems, an interaction may additionally contain stable attributes such as category, brand, and seller; dynamic states such as price, inventory, and promotion status; and interaction-level context such as behavior type and timestamp. In particular, dynamic states and interaction context are tied to the moment at which an interaction occurs and therefore cannot be reliably encoded into a statically assigned SID.

A straightforward solution is to represent each additional field as an individual token and feed all field tokens into the generative backbone. Although this preserves rich information, it substantially expands the input context. For a history containing $L$ interactions, representing each interaction with multiple field tokens increases the effective sequence length in proportion to the number of fields. Since the computational and memory costs of self-attention grow approximately quadratically with sequence length~\cite{Transformer,SETRec,RPG,Grace}, such expansion incurs considerable training and inference overhead and reduces the number of historical interactions that the model can process. Existing methods that append attribute, behavior, or temporal tokens face the same scalability issue as the number of information dimensions increases~\cite{MBSR,Grace,TAFAME,ACERec}.

Another more efficient alternative is to aggregate the SID, item attributes, dynamic features, and interaction context of each behavior into a single compact representation before feeding it into the generative backbone. However, directly compressing heterogeneous fields before sufficient interaction creates a premature information bottleneck. Especially, field importance depends not only on the field content itself but also on the surrounding behavioral sequence. As illustrated in Figure~\ref{fig:intro_1}, brand information becomes highly predictive when a user repeatedly interacts with items from the same brand. In contrast, when a user shows stronger purchase intent for higher-priced items while only clicking or collecting cheaper ones, price and behavior type become more informative than brand. Consequently, the information retained during compression should be conditioned on the current sequential context.

This creates a fundamental trade-off. Fully expanding heterogeneous fields and delegating their interaction to a large generative backbone preserves information but incurs prohibitive context and computational costs. In contrast, directly compressing them without sufficient interaction is efficient but may irreversibly discard context-dependent signals. The key challenge is therefore to contextualize multidimensional behavioral information at low cost before compression, while ensuring that each historical interaction ultimately occupies only one token in the backbone input.

To address this challenge, we propose \textbf{SPARC}, a \uline{\textbf{S}}equence-aware \uline{\textbf{P}}rogressive \uline{\textbf{A}}ttribute \uline{\textbf{R}}outing and \uline{\textbf{C}}ompression Framework for Generative Recommendation. Rather than reconstructing the SIDs of target items, SPARC focuses on enriching the historical inputs of the generative model. It employs a lightweight front-end module to contextualize and progressively compress heterogeneous behavioral information, ensuring that each historical interaction is represented by a single backbone token. SPARC therefore preserves richer behavioral signals without increasing the input length of the generative backbone.

SPARC consists of three stages. First, the \textit{Field-wise Context Modeling} module organizes representations of the same field type along the user history and models their sequential evolution, producing a context-aware representation for each field. Second, the \textit{Context-aware Attribute Routing} module jointly considers the original field representation, its contextualized representation, and its field identity, and routes heterogeneous fields into a fixed number of intermediate tokens to retain complementary information under a constrained representation budget. Finally, the \textit{Sequence-level Token Consolidation} module reorganizes the intermediate tokens of all historical interactions into a fine-grained sequence and performs lightweight cross-interaction modeling before integrating the tokens of each interaction into a single compact representation. Following the principle of \textbf{interacting before compressing}, SPARC provides the generative backbone with more informative historical representations without increasing its input length.

Our main contributions are summarized as follows:
\begin{itemize}[leftmargin=*]
    \vspace{5pt}
    \item We identify a fundamental representation challenge in industrial generative recommendation: statically assigned SIDs cannot fully capture dynamic item states and interaction-specific context, whereas explicitly tokenizing such information causes substantial context expansion.
    \vspace{5pt}
    \item We propose SPARC, a sequence-aware progressive attribute routing and compression framework that contextualizes heterogeneous behavioral signals before compression and preserves context-relevant information within a fixed item-level token budget.
    \vspace{5pt}
    \item We conduct extensive experiments on an industrial Taobao dataset and two public Amazon datasets. The results demonstrate that SPARC consistently outperforms strong baselines, while further comparisons with static compression variants and routing analyses verify the advantage of context-conditioned information retention.
\end{itemize}

\section{Preliminary}
In this section, we introduce the basic formulation of SID-based generative recommendation and then formalize the multi-field historical compression problem studied in this work.

\subsection{Generative Recommendation}

Generative recommendation formulates item recommendation as an autoregressive generation problem over discrete item tokens~\cite{TIGER,LETTER,Cost,ETEGRec}. Given an item set $\mathcal{I}$, an item encoder first maps each item $i \in \mathcal{I}$ into a continuous representation $\mathbf{h}_i$ by using item content, structured attributes, multimodal features, or collaborative signals. An item tokenizer then discretizes $\mathbf{h}_i$ into a sequence of SIDs:
\begin{equation}
\mathbf{s}_i = (s_i^1, s_i^2, \ldots, s_i^K),
\end{equation}
where $K$ is the SID length and each $s_i^k$ is selected from a discrete codebook. Existing methods obtain such discrete tokens through clustering, vector quantization, residual quantization, or learnable tokenization objectives~\cite{VQRec,TIGER,LETTER,Cost,ETEGRec}.

Given a user behavior sequence $\mathcal{S}_u=(i_1,i_2,\ldots,i_L)$, the generative recommender converts each historical item into its SID sequence and feeds the resulting token sequence into a Transformer-based backbone. The model predicts the SID of the next item $i_{L+1}$ in an autoregressive manner:
\begin{equation}
p_{\theta}(\mathbf{s}_{i_{L+1}} \mid \mathcal{S}_u)
=
\prod_{k=1}^{K}
p_{\theta}
\left(
s_{i_{L+1}}^k
\mid
\mathbf{s}_{i_1},\ldots,\mathbf{s}_{i_L},
s_{i_{L+1}}^{<k}
\right),
\end{equation}
where $\theta$ denotes the parameters of the generative recommender, and $s_{i_{L+1}}^{<k}$ denotes the previously generated target SID tokens before step $k$. The model is optimized over all users by minimizing the negative log-likelihood of the target SID sequence:
\begin{equation}
\mathcal{L}_{\mathrm{rec}}
=
-\sum_{u}
\sum_{k=1}^{K}
\log
p_{\theta}
\left(
s_{i_{L+1}}^k
\mid
\mathbf{s}_{i_1},\ldots,\mathbf{s}_{i_L},
s_{i_{L+1}}^{<k}
\right).
\end{equation}

In this paradigm, SIDs serve as discrete substitutes for conventional dense item ID features. Specifically, they provide the primary identity signal utilized by the generative recommender to reference historical and target items, thereby remaining crucial for preserving stable item identity during user sequence modeling.

\subsection{Multi-field Historical Compression}

In industrial recommendation scenarios, each historical interaction usually contains multiple heterogeneous fields beyond the item SID. For the $t$-th interaction, we denote its field representations as
\begin{equation}
\mathbf{E}_t =
[\mathbf{e}_t^1,\mathbf{e}_t^2,\ldots,\mathbf{e}_t^F]
\in \mathbb{R}^{F \times d},
\end{equation}
where $F$ is the number of fields, $d$ is the representation dimension, and $\mathbf{e}_t^f$ is the representation of the $f$-th field. These fields may include SID tokens, category, brand, seller, price, behavior type, timestamp, and other side information. If a field contains multiple raw tokens, we obtain $\mathbf{e}_t^f$ through token embedding followed by mean pooling or another lightweight field encoder.

A straightforward way to use all information is to feed all field tokens of all historical interactions into the generative backbone:
\begin{equation}
\mathbf{H}_u^{\mathrm{full}}
=
[
\mathbf{E}_1,\mathbf{E}_2,\ldots,\mathbf{E}_L
].
\end{equation}
This representation preserves fine-grained field information and allows the backbone to model field-level relations. However, it expands the input length from $L$ item-level representations to approximately $LF$ field-level representations, which substantially increases the computational and memory costs of Transformer-based generative recommenders.

Therefore, we study item-level compression for heterogeneous historical interactions. Let $\mathbf{E}_{1:L}=\{\mathbf{E}_1,\mathbf{E}_2,\ldots,\mathbf{E}_L\}$ denote the field representations of the whole user history. The goal is to compress each multi-field interaction into a compact representation before feeding it into the generative backbone:
\begin{equation}
\mathbf{z}_t
=
\mathcal{C}_{\psi}
\left(
\mathbf{E}_t,\mathbf{E}_{1:L}
\right)
\in \mathbb{R}^{d},
\end{equation}
where $\mathcal{C}_{\psi}$ is the compression function parameterized by $\psi$. The compressed user sequence is then written as
\begin{equation}
\mathbf{H}_u^{\mathrm{comp}}
=
[\mathbf{z}_1,\mathbf{z}_2,\ldots,\mathbf{z}_L],
\end{equation}
which keeps the same length as a conventional item-level history while allowing each token to contain richer multi-field information.

Existing item-wise compression methods (e.g., mean pooling or MLP-based compression) usually adopt a statically context-independent form:
\begin{equation}
\mathbf{z}_t
=
\mathcal{C}_{\psi}^{\mathrm{static}}(\mathbf{E}_t),
\end{equation}
where the retained information only depends on the fields of the current interaction. In contrast, this work focuses on dynamically context-conditioned compression:
\begin{equation}
\mathbf{z}_t
=
\mathcal{C}_{\psi}^{\mathrm{context}}(\mathbf{E}_t,\mathbf{E}_{1:L}),
\end{equation}
where the compression of each interaction can be adjusted according to the surrounding behavior sequence. Under this formulation, the key problem is how to preserve stable item identity from SIDs while selectively retaining useful side information conditioned on the sequential context.

\section{Method}
\begin{figure*}[t]
\centering
\includegraphics[width=\textwidth]{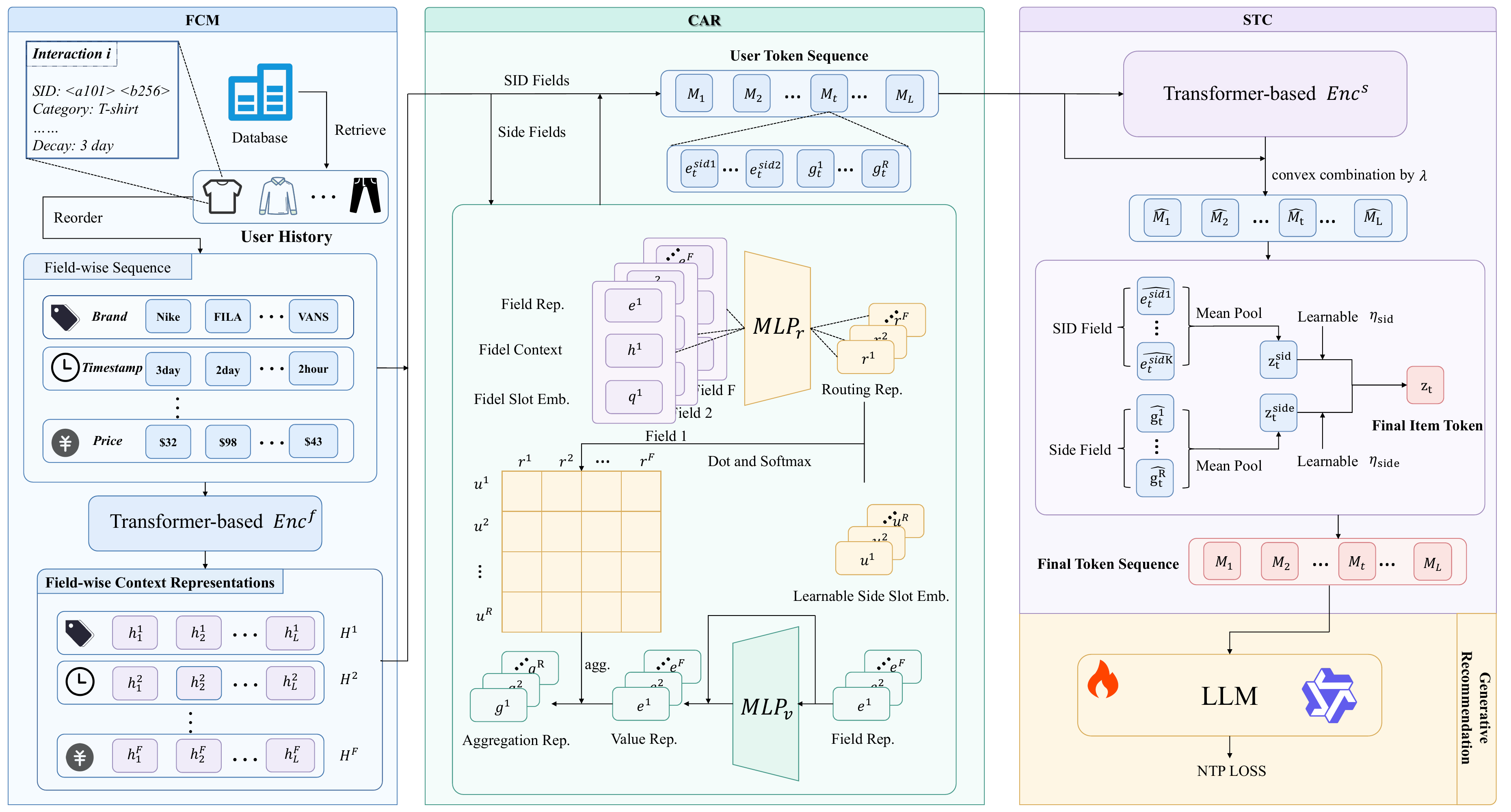}
\caption{Overall framework of SPARC. SPARC compresses each multi-field historical interaction into a single backbone token through three progressive stages. FCM models field-wise sequential context, CAR preserves SID-based item identity and routes context-conditioned side information into intermediate tokens, and STC refines and consolidates these intermediate tokens into final historical tokens. The compressed sequence is then fed into the generative recommendation backbone for autoregressive target SID prediction without increasing the backbone input length.}
\label{fig:main}
\end{figure*}

In this section, we introduce the proposed SPARC framework. We first present the overall architecture, then detail its three key modules, and finally describe the training procedure.

\subsection{Overall Framework}

SPARC is a lightweight context-conditioned compressor placed before the generative recommendation backbone, as illustrated in Figure~\ref{fig:main}. Given the multi-field representations of a user history, SPARC first models the sequential context of each field type through Field-wise Context Modeling (FCM), then preserves SID-based item identity and routes contextual side information into intermediate tokens through Context-aware Attribute Routing (CAR), and finally consolidates these intermediate tokens into one compact token for each historical interaction through Sequence-level Token Consolidation (STC). In this way, SPARC enriches historical item representations with context-dependent side information while keeping the backbone input length unchanged and leaving the target SID generation objective intact.

\subsection{Field-wise Context Modeling (FCM)}

FCM captures field-specific sequential patterns before item-level compression. Given the historical field representations $\mathbf{E}_{1:L}$, SPARC first groups representations with the same field type along the user history. For the $f$-th field, the field-wise sequence is defined as
\begin{equation}
\mathbf{X}^{f}
=
[\mathbf{e}_1^f,\mathbf{e}_2^f,\ldots,\mathbf{e}_L^f].
\label{eq:fcm_sequence}
\end{equation}
A lightweight sequence encoder $\mathrm{Enc}^{\mathrm{f}}$ is then applied to each field-wise sequence:
\begin{equation}
\mathbf{H}^{f}
=
[\mathbf{h}_1^f,\mathbf{h}_2^f,\ldots,\mathbf{h}_L^f]
=
\mathrm{Enc}^{\mathrm{f}}(\mathbf{X}^{f}),
\label{eq:fcm_encoding}
\end{equation}
where $\mathbf{X}^{f}$ denotes the length-$L$ sequence formed by the $f$-th field across the user history, $\mathrm{Enc}^{\mathrm{f}}(\cdot)$ is a lightweight field-wise sequence encoder, and $\mathbf{H}^{f}$ is the resulting contextualized sequence. Specifically, $\mathbf{h}_t^f$ denotes the contextualized representation of field $f$ at position $t$.

This design introduces sequence context before heterogeneous fields are compressed. Instead of directly mixing different fields within each interaction, FCM first models how each field type evolves along the user history, enabling subsequent routing to consider both the original field representation $\mathbf{e}_t^f$ and its context-aware counterpart $\mathbf{h}_t^f$.
\subsection{Context-aware Attribute Routing (CAR)}

CAR compresses the heterogeneous fields of each interaction into a small set of intermediate tokens. We divide the fields into two disjoint groups: SID fields $\mathcal{F}_{\mathrm{sid}}$ and side fields $\mathcal{F}_{\mathrm{side}}$. Since SID tokens provide the primary item identity signal in generative recommendation, SPARC explicitly preserves them instead of mixing them with side fields during routing. For each interaction, the identity tokens are constructed as
\begin{equation}
\mathbf{I}_t
=
[\mathbf{e}_t^f]_{f\in\mathcal{F}_{\mathrm{sid}}}.
\label{eq:car_identity}
\end{equation}
where $\mathbf{I}_t$ denotes the preserved SID-based identity token set of interaction $t$.

For side fields, CAR performs context-conditioned routing. For each side field $f\in\mathcal{F}_{\mathrm{side}}$, we build a routing representation by combining its original representation, contextualized representation, and field embedding:
\begin{equation}
\mathbf{r}_t^f
=
\mathrm{MLP}_{\mathrm{r}}
\left(
[\mathbf{e}_t^f;\mathbf{h}_t^f;\mathbf{q}^f]
\right),
\label{eq:car_routing_repr}
\end{equation}
where $[\cdot;\cdot]$ denotes concatenation, $\mathrm{MLP}_{\mathrm{r}}(\cdot)$ is the routing network, and $\mathbf{q}^f$ is a learnable embedding that encodes the identity of field $f$. Given $R$ learnable side slots $\{\mathbf{u}_1,\ldots,\mathbf{u}_R\}$, the routing weight from side field $f$ to slot $r$ is computed by normalizing over all side fields:
\begin{equation}
\alpha_{t,r}^{f}
=
\frac{
\exp\left((\mathbf{r}_t^f)^{\top}\mathbf{u}_r\right)
}{
\sum_{g\in\mathcal{F}_{\mathrm{side}}}
\exp\left((\mathbf{r}_t^g)^{\top}\mathbf{u}_r\right)
}.
\label{eq:car_routing_weight}
\end{equation}
where $\alpha_{t,r}^{f}$ measures the contribution of side field $f$ to the $r$-th side slot at interaction position $t$, and $g$ is the summation index over side fields. Since the normalization is performed over $\mathcal{F}_{\mathrm{side}}$, each side slot adaptively selects information from different side fields under the current sequence context.

The $r$-th side token is obtained by aggregating side-field values:
\begin{equation}
\mathbf{g}_t^r
=
\sum_{f\in\mathcal{F}_{\mathrm{side}}}
\alpha_{t,r}^{f}\mathbf{v}_t^f,
\label{eq:car_side_token}
\end{equation}
where $\mathbf{g}_t^r$ denotes the $r$-th routed side token and $\mathbf{v}_t^f$ is the value representation of side field $f$. The value representation is defined by a residual value projection:
\begin{equation}
\mathbf{v}_t^f
=
\mathbf{e}_t^f
+
\mathrm{MLP}_{\mathrm{v}}(\mathbf{e}_t^f),
\label{eq:car_value}
\end{equation}
where $\mathrm{MLP}_{\mathrm{v}}(\cdot)$ is a lightweight value projection network. This residual design keeps the side-token values close to the original field representations, so that CAR mainly learns how to route side information rather than rewriting the field space from scratch.

The output of CAR is the intermediate token set:
\begin{equation}
\mathbf{M}_t
=
[\mathbf{I}_t;\mathbf{g}_t^1,\ldots,\mathbf{g}_t^R].
\label{eq:car_output}
\end{equation}
where $\mathbf{M}_t$ denotes the intermediate token set of interaction $t$, consisting of the preserved SID identity tokens and $R$ routed side tokens. Therefore, CAR preserves stable item identity through SID tokens and uses sequence-aware routing to retain complementary side information under a limited token budget.
\subsection{Sequence-level Token Consolidation (STC)}

After CAR, each interaction is represented by multiple intermediate tokens. STC further models their cross-interaction dependency and consolidates them into a single backbone token. We first concatenate the intermediate tokens of all historical interactions into a fine-grained token sequence:
\begin{equation}
\mathbf{M}_{1:L}
=
[\mathbf{M}_1,\mathbf{M}_2,\ldots,\mathbf{M}_L].
\label{eq:stc_flatten}
\end{equation}
where $\mathbf{M}_{1:L}$ denotes the ordered sequence formed by all intermediate token sets in the user history. Although these intermediate tokens are temporarily flattened for sequence-level interaction, their interaction boundaries are preserved for the final within-interaction consolidation.

A lightweight sequence encoder $Enc^s$ is then applied to the flattened sequence:
\begin{equation}
\widetilde{\mathbf{M}}_{1:L}
=
\mathrm{Enc}^{\mathrm{s}}(\mathbf{M}_{1:L}),
\label{eq:stc_encoding}
\end{equation}
where $\mathrm{Enc}^{\mathrm{s}}(\cdot)$ is a lightweight sequence-level encoder and $\widetilde{\mathbf{M}}_{1:L}$ denotes the sequence-enhanced intermediate tokens. Positional embeddings and token-type embeddings are used to distinguish interaction positions and intermediate token roles.

To avoid unstable changes to the compressed token space, SPARC introduces sequence-level interaction through a gated residual update:
\begin{equation}
\widehat{\mathbf{M}}_{1:L}
=
\mathbf{M}_{1:L}
+
\lambda
\left(
\widetilde{\mathbf{M}}_{1:L}
-
\mathbf{M}_{1:L}
\right),
\label{eq:stc_residual}
\end{equation}
where $\widehat{\mathbf{M}}_{1:L}$ denotes the updated intermediate tokens and $\lambda$ is a learnable residual gate. This design allows STC to gradually incorporate cross-interaction information while preserving the original intermediate representations.

Finally, STC summarizes identity tokens and side tokens separately:
\begin{equation}
\mathbf{z}_t^{\mathrm{id}}
=
\mathrm{Pool}_{\mathrm{id}}(\widehat{\mathbf{M}}_t),
\quad
\mathbf{z}_t^{\mathrm{side}}
=
\mathrm{Pool}_{\mathrm{side}}(\widehat{\mathbf{M}}_t),
\label{eq:stc_summary}
\end{equation}
where $\widehat{\mathbf{M}}_t$ denotes the updated intermediate token set corresponding to interaction $t$, and $\mathrm{Pool}_{\mathrm{id}}(\cdot)$ and $\mathrm{Pool}_{\mathrm{side}}(\cdot)$ summarize the updated identity tokens and side tokens within the same interaction, respectively. STC then fuses them into the final historical token:
\begin{equation}
\mathbf{z}_t
=
\eta_{\mathrm{id}}\mathbf{z}_t^{\mathrm{id}}
+
\eta_{\mathrm{side}}\mathbf{z}_t^{\mathrm{side}},
\label{eq:stc_fusion}
\end{equation}
where $\eta_{\mathrm{id}}$ and $\eta_{\mathrm{side}}$ are learnable fusion weights. We denote the final compressed historical sequence as $\mathbf{Z}_{1:L}=[\mathbf{z}_1,\ldots,\mathbf{z}_L]$, which keeps the same length as the original item-level history, while each token contains both preserved SID identity and context-conditioned side information.

\subsection{Training Pipeline}
\label{train_pipeline}
The training procedure of SPARC is summarized in Algorithm~\ref{alg:sparc}. SPARC is optimized jointly with the generative recommendation backbone, while the item tokenizer and the target SID space remain unchanged. For each mini-batch, SPARC first transforms the multi-field historical inputs into context-conditioned item-level tokens, which are then used by the backbone to predict the target SID sequence. Here, $y$ denotes the target item, $\mathbf{s}_{y}=(s_y^1,\ldots,s_y^K)$ is its SID sequence, $\mathbf{s}_{y}^{<k}$ denotes the previously generated target SID tokens before step $k$, $\psi$ denotes the parameters of SPARC, and $\theta$ denotes the parameters of the generative backbone.

\begin{algorithm}[t]
\caption{Training Pipeline of SPARC}
\label{alg:sparc}
\begin{algorithmic}[1]
\Require Historical fields $\mathbf{E}_{1:L}$, target SIDs $\mathbf{s}_{y}=(s_y^1,\ldots,s_y^K)$, parameters $\psi,\theta$
\Ensure Optimized $\psi,\theta$

\For{each mini-batch}

    \Statex $\triangleright$ \textbf{FCM}
    \State Build $\{\mathbf{X}^f\}_{f=1}^{F}$ by Eq.~\eqref{eq:fcm_sequence}.
    \State Encode $\{\mathbf{H}^f\}_{f=1}^{F}$ by Eq.~\eqref{eq:fcm_encoding}.

    \Statex $\triangleright$ \textbf{CAR}
    \State Preserve $\mathbf{I}_{1:L}$ by Eq.~\eqref{eq:car_identity}.
    \State Compute $\mathbf{r}_t^f$ and $\alpha_{t,r}^{f}$ by Eqs.~\eqref{eq:car_routing_repr}--\eqref{eq:car_routing_weight}.
    \State Compute $\mathbf{v}_t^f$ and $\mathbf{g}_t^r$ by Eqs.~\eqref{eq:car_value}--\eqref{eq:car_side_token}.
    \State Form $\mathbf{M}_{1:L}$ by Eq.~\eqref{eq:car_output}.

    \Statex $\triangleright$ \textbf{STC}
    \State Flatten $\mathbf{M}_{1:L}$ by Eq.~\eqref{eq:stc_flatten}.
    \State Encode $\widetilde{\mathbf{M}}_{1:L}$ by Eq.~\eqref{eq:stc_encoding}.
    \State Update $\widehat{\mathbf{M}}_{1:L}$ by Eq.~\eqref{eq:stc_residual}.
    \State Fuse $\mathbf{Z}_{1:L}$ by Eqs.~\eqref{eq:stc_summary}--\eqref{eq:stc_fusion}.

    \Statex $\triangleright$ \textbf{Generation}
    \State Predict $p_{\theta}(s_y^k \mid \mathbf{Z}_{1:L}, \mathbf{s}_{y}^{<k})$ for $k=1,\ldots,K$.
    \State $\mathcal{L}\leftarrow-\sum_{k=1}^{K}\log p_{\theta}(s_y^k \mid \mathbf{Z}_{1:L}, \mathbf{s}_{y}^{<k})$.
    \State Optimize $\psi,\theta$ w.r.t. $\mathcal{L}$.

\EndFor
\end{algorithmic}
\end{algorithm}

Specifically, for each mini-batch, FCM first reorganizes the historical fields by field type and encodes each field-wise sequence to obtain contextualized field representations. CAR then preserves SID fields as identity tokens and routes side fields into intermediate tokens according to their context-aware routing weights. STC further models the intermediate-token sequence and fuses the identity and side summaries into the compressed historical sequence $\mathbf{Z}_{1:L}$. Finally, the generative backbone predicts the target SID tokens conditioned on $\mathbf{Z}_{1:L}$, and the parameters of SPARC and the backbone are jointly updated by minimizing the autoregressive recommendation loss. Therefore, SPARC only changes the representation of historical interactions and does not alter the target SID generation objective.

\section{Experiment}
In this section, we conduct a series of experiments to answer the following research questions: \\
\textbf{RQ1}: How does the proposed SPARC framework perform compared to existing generative recommendation methods? \\
\textbf{RQ2}: Is context-conditioned compression more effective than static compression strategies for multi-field historical modeling? \\
\textbf{RQ3}: Does the slots of CIR learn division of labor? \\
\textbf{RQ4}: Do the same item's CIR slot weights actually differ across different contexts?

\subsection{Experimental Setting}
\subsubsection{Datasets.}
For brevity, details of the datasets are deferred to Section~\ref{dataset}.

\subsubsection{Baselines.}
For brevity, details of the baselines are deferred to Section~\ref{baselines}.

\subsubsection{Evaluation Metrics.}

For the industrial \emph{TaoBao} dataset, following prior work~\cite{RankGR}, we adopt Hit Rate (HR) as the main evaluation metric for retrieval performance. Given the top-$K$ retrieved item set $\mathcal{I}_u^K$ and the ground-truth interacted item set $\mathcal{I}_u^{\mathrm{truth}}$ for user $u$, HR@K is defined as
\begin{equation}
\mathrm{HR}@K
=
\frac{1}{|\mathcal{U}|}
\sum_{u\in\mathcal{U}}
\frac{
|\mathcal{I}_u^K \cap \mathcal{I}_u^{\mathrm{truth}}|
}{
|\mathcal{I}_u^{\mathrm{truth}}|
},
\end{equation}
where $\mathcal{U}$ denotes the user set. To evaluate the model under different behavior-level feedback signals, we report HR for click and pageview behaviors on the \emph{TaoBao} dataset, denoted as $\mathrm{HR}^{\mathrm{Click}}@K$ and $\mathrm{HR}^{\mathrm{PV}}@K$, respectively. In our experiments, we report results at $K=20$ and $K=1000$.

For the public datasets, we also adopt HR@K as evaluation metrics. And We report HR@20, HR@50, HR@100, and HR@500 on the public datasets.
\subsubsection{Implementation Details.}

For all baseline methods, we follow the implementation and experimental settings of RankGR~\cite{RankGR} to ensure a fair comparison. For SPARC, we apply the proposed framework on top of RankGR, using RankGR as the generative recommendation backbone. 

For SPARC-specific hyperparameters, each historical interaction is represented by 9 fields, consisting of 2 SID fields and 7 side fields. The 2 SID fields are directly preserved as identity tokens, while the 7 side fields are routed into $R=2$ side tokens. As a result, each interaction is represented by 4 intermediate tokens before sequence-level consolidation. Both the field-wise context encoder in FCM and the sequence-level encoder in STC are implemented as 2-layer Transformer encoders. To stabilize training, the sequence-level residual gate is initialized with a logit value of $-5$, allowing the STC refinement to be introduced gradually at the early stage of optimization. The fusion weights for the identity and side summaries are initialized equally. Unless otherwise specified, the same SPARC hyperparameters are used across all datasets.

\subsection{Overall Performance (RQ1)}
\begin{table*}[]
\centering
\caption{Overall performance comparison between baselines and our proposed SPARC. Specifically, SPARC is applied on top of RankGR. The best and second-best results are highlighted in bold and underlined, respectively.}
\label{tab:main}
\resizebox{\textwidth}{!}{%
\begin{tabular}{cccccccclccccc}
\hline
 &                       & \multicolumn{6}{c}{Conventional Method}             &  & \multicolumn{5}{c}{Generative Method}                                             \\ \cline{3-8} \cline{10-14} 
\multirow{-2}{*}{Dataset} &
  \multirow{-2}{*}{Metric} &
  YouTubeDNN &
  SASRec &
  BERT4Rec &
  Caser &
  NextItNet &
  CORE &
   &
  HSTU &
  TIGER &
  FORGE &
  RankGR &
  \cellcolor[HTML]{EFEFEF}SPARC \\ \hline
 & $HR^{click}@20$   & 0.0337 & 0.0496 & 0.0503 & 0.0429 & 0.0487 & 0.0557 &  & 0.0570 & 0.1008 & 0.1283 & {\ul 0.1568} & \cellcolor[HTML]{EFEFEF}\textbf{0.1669} \\
 & $HR^{click}@1000$ & 0.1164 & 0.1670 & 0.1676 & 0.1567 & 0.1644 & 0.2012 &  & 0.2022 & 0.3604 & 0.4666 & {\ul 0.5777} & \cellcolor[HTML]{EFEFEF}\textbf{0.5883} \\
 & $HR^{PV}@20$      & 0.0149 & 0.0218 & 0.0181 & 0.0167 & 0.0172 & 0.0224 &  & 0.0323 & 0.1023 & 0.1335 & {\ul 0.1562} & \cellcolor[HTML]{EFEFEF}\textbf{0.1670} \\
\multirow{-4}{*}{TaoBao} &
  $HR^{PV}@1000$ &
  0.1183 &
  0.2160 &
  0.2266 &
  0.2039 &
  0.2071 &
  0.2309 &
   &
  0.2490 &
  0.4171 &
  0.4896 &
  {\ul 0.6228} &
  \cellcolor[HTML]{EFEFEF}\textbf{0.6319} \\ \hline
 & $HR@20$           & 0.0357 & 0.0372 & 0.0370 & 0.0355 & 0.0351 & 0.0302 &  & 0.0374 & 0.0209 & 0.0435 & {\ul 0.0466} & \cellcolor[HTML]{EFEFEF}\textbf{0.0794} \\
 & $HR@50$           & 0.0671 & 0.0706 & 0.0711 & 0.0664 & 0.0661 & 0.0669 &  & 0.0714 & 0.0496 & 0.0761 & {\ul 0.0784} & \cellcolor[HTML]{EFEFEF}\textbf{0.1311} \\
 & $HR@100$          & 0.0919 & 0.0976 & 0.0979 & 0.0909 & 0.0902 & 0.0919 &  & 0.0993 & 0.0590 & 0.1088 & {\ul 0.1117} & \cellcolor[HTML]{EFEFEF}\textbf{0.1843} \\
\multirow{-4}{*}{Beauty} &
  $HR@500$ &
  0.1967 &
  0.2119 &
  0.2121 &
  0.1953 &
  0.1956 &
  0.2135 &
   &
  0.2167 &
  0.1312 &
  0.2250 &
  {\ul 0.2289} &
  \cellcolor[HTML]{EFEFEF}\textbf{0.3571} \\ \hline
 & $HR@20$           & 0.0340 & 0.0353 & 0.0351 & 0.0338 & 0.0334 & 0.0287 &  & 0.0355 & 0.0199 & 0.0414 & {\ul 0.0443} & \cellcolor[HTML]{EFEFEF}\textbf{0.0743} \\
 & $HR@50$           & 0.0629 & 0.0662 & 0.0667 & 0.0622 & 0.0620 & 0.0627 &  & 0.0669 & 0.0465 & 0.0714 & {\ul 0.0735} & \cellcolor[HTML]{EFEFEF}\textbf{0.1279} \\
 & $HR@100$          & 0.0834 & 0.0886 & 0.0888 & 0.0826 & 0.0819 & 0.0834 &  & 0.0901 & 0.0535 & 0.0988 & {\ul 0.1014} & \cellcolor[HTML]{EFEFEF}\textbf{0.1795} \\
\multirow{-4}{*}{Toys} &
  $HR@500$ &
  0.1886 &
  0.2032 &
  0.2034 &
  0.1873 &
  0.1875 &
  0.2047 &
   &
  0.2078 &
  0.1258 &
  0.2158 &
  {\ul 0.2195} &
  \cellcolor[HTML]{EFEFEF}\textbf{0.3605} \\ \hline
\end{tabular}%
}
\end{table*}

We evaluate the overall performance of the compared approaches across all three datasets. The quantitative results are summarized in Table~\ref{tab:main}, from which several observations can be drawn:
\begin{itemize}[leftmargin=*]
    \vspace{5pt}
    \item \textbf{SPARC consistently achieves the best performance.}
    SPARC outperforms all baselines on both the industrial \emph{TaoBao} dataset and the public Amazon datasets, demonstrating the effectiveness of context-conditioned historical compression. The improvement on \emph{TaoBao} is relatively moderate, likely because its massive interaction scale already alleviates part of the representation bottleneck, whereas the sparser public datasets benefit more from adaptive information retention.
    \vspace{5pt}
    \item \textbf{SPARC improves over the RankGR backbone.}
    Since SPARC is built upon RankGR, the consistent gains over RankGR directly verify the value of our proposed compression framework. This shows that better historical interaction representation can further enhance generative retrieval without changing the backbone generation paradigm.
    \vspace{5pt}
    \item \textbf{Generative methods generally outperform conventional methods.}
    Compared with conventional recommenders, generative methods usually achieve stronger performance, especially at larger cutoffs. This is mainly because SID-based generation introduces structured item tokens that encode prior semantic or collaborative information, improving long-tail generalization and enabling more global token-level discrimination.
\end{itemize}

\subsection{Ablation Study (RQ2)}

\begin{table}[]
\centering
\caption{Ablation study of SPARC on the \emph{TaoBao} dataset.}
\label{tab:ablation}
\resizebox{\columnwidth}{!}{%
\begin{tabular}{ccccc}
\hline
Method & $HR^{click}@20$ & $HR^{click}@1000$ & $HR^{PV}@20$ & $HR^{PV}@1000$ \\ \hline
SPARC & \textbf{0.1669} & \textbf{0.5883} & \textbf{0.1670} & \textbf{0.6319} \\ \hline
RankGR & 0.1568 & 0.5777 & 0.1562 & 0.6228 \\
\textit{w/ MLP} & 0.1576 & 0.5796 & 0.1604 & 0.6269 \\
\textit{w/ QFormer} & {\ul 0.1609} & {\ul 0.5835} & {\ul 0.1633} & {\ul 0.6302} \\
\textit{w/ Modulated} & 0.1560 & 0.5762 & 0.1585 & 0.6234 \\ \hline
\end{tabular}%
}
\end{table}

To validate the design rationale behind SPARC, we compare it with RankGR and several controlled variants built upon the same generative backbone. These variants introduce different static compression strategies for multi-field historical interactions, allowing us to examine whether a stronger static compressor is sufficient for improving historical representation learning.

Specifically, we introduce the following variants:

\begin{itemize}[leftmargin=*]
    \vspace{5pt}
    \item \textbf{w/ MLP}: This variant concatenates the field representations of each interaction and transforms them through an MLP-based compressor into a single historical token.
    \vspace{5pt}
    \item \textbf{w/ QFormer}: This variant adopts a lightweight query-based Transformer compressor, where learnable query tokens interact with the field representations via cross-attention and produce the compressed token.
    \vspace{5pt}
    \item \textbf{w/ Modulated}: This variant projects each content field into a dedicated subspace and leverages the context fields (e.g., recency and behavior type) of the interaction to modulate the concatenated content representation via learned scaling factors before projecting it into the compressed token.
\end{itemize}

Table~\ref{tab:ablation} reports the results on \emph{TaoBao}, from which we draw the following observations:
\begin{itemize}[leftmargin=*]
    \vspace{5pt}
    \item \textbf{Static compression provides limited improvements.}
    Compared with RankGR, some static variants achieve better performance, indicating that improving the compression function can help multi-field historical modeling. However, the gains are relatively small and not consistent across all designs, suggesting that compressor capacity alone is not the key factor.
    \vspace{5pt}
    \item \textbf{More complex compression does not guarantee better results.}
    Among the static variants, QFormer performs the best, while MLP and Modulated bring only marginal gains or even slight degradation on some metrics. This shows that directly transforming multi-field representations with a stronger static compressor may introduce unnecessary distortion when the compression process is not guided by sequential context.
    \vspace{5pt}
    \item \textbf{SPARC consistently outperforms all static variants.}
    SPARC achieves the best performance across all metrics. Since all variants are built upon the same generative backbone, the superiority of SPARC verifies the effectiveness of context-conditioned information retention, rather than merely increasing the expressiveness of the compression module.
\end{itemize}

\subsection{In-depth Analysis (RQ3 \& RQ4)}

To further interpret how SPARC performs context-conditioned information retention, we analyze the routing weights learned by CAR. Since SID fields are explicitly preserved as identity tokens and are not involved in side-field routing, we focus on the routing distributions over side fields. Specifically, we conduct two analyses: a slot-level routing analysis to examine whether different side slots learn differentiated field preferences, and a same-item case study to examine whether the routing distribution changes under different user contexts.

\textbf{Slot-level Routing Pattern.}
\begin{figure}[t]
\centering
\includegraphics[width=\linewidth]{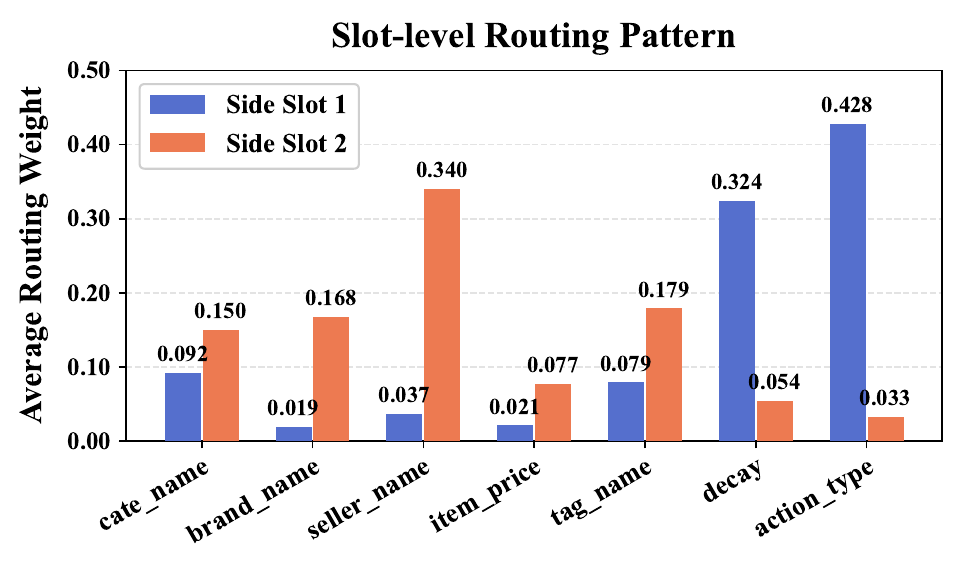}
\caption{Slot-level routing patterns learned by CAR. The values denote the average routing weights assigned by each side slot to different side fields over the evaluation set. SID identity slots are omitted since they are directly preserved rather than routed.}
\label{fig:slot_weight}
\end{figure}
We analyze the average routing weights assigned by different side slots to each side field. Specifically, for each side slot, we average its routing weights over all interactions in the evaluation set, obtaining a field-level routing distribution for each slot.

The result is visualized in Figure~\ref{fig:slot_weight}, from which we can draw the following observations:
\begin{itemize}[leftmargin=*]
    \vspace{5pt}
    \item \textbf{The two side slots learn clearly different routing patterns.}
    Side Slot 1 assigns much larger weights to \emph{action\_type} and \emph{decay}, while Side Slot 2 mainly focuses on \emph{seller\_name} and also assigns relatively higher weights to \emph{brand\_name}, \emph{tag\_name}, and \emph{cate\_name}. This indicates that the two learnable side slots do not collapse into redundant copies. Instead, they develop a meaningful division of labor for retaining complementary side information.

    \vspace{5pt}
    \item \textbf{Context-related and seller-related fields receive dominant routing weights.}
    Among all side fields, \emph{action\_type}, \emph{decay}, and \emph{seller\_name} receive the most prominent routing weights. This suggests that CAR tends to preserve behavior-context signals and seller-related information during compression. Such a pattern is consistent with the intuition that behavior type and recency describe the current interaction context, while seller information provides an important attribute signal for distinguishing items in industrial recommendation scenarios.
\end{itemize}

\textbf{Context-dependent Routing for the Same Item.}
\begin{figure}[t]
\centering
\includegraphics[width=\linewidth]{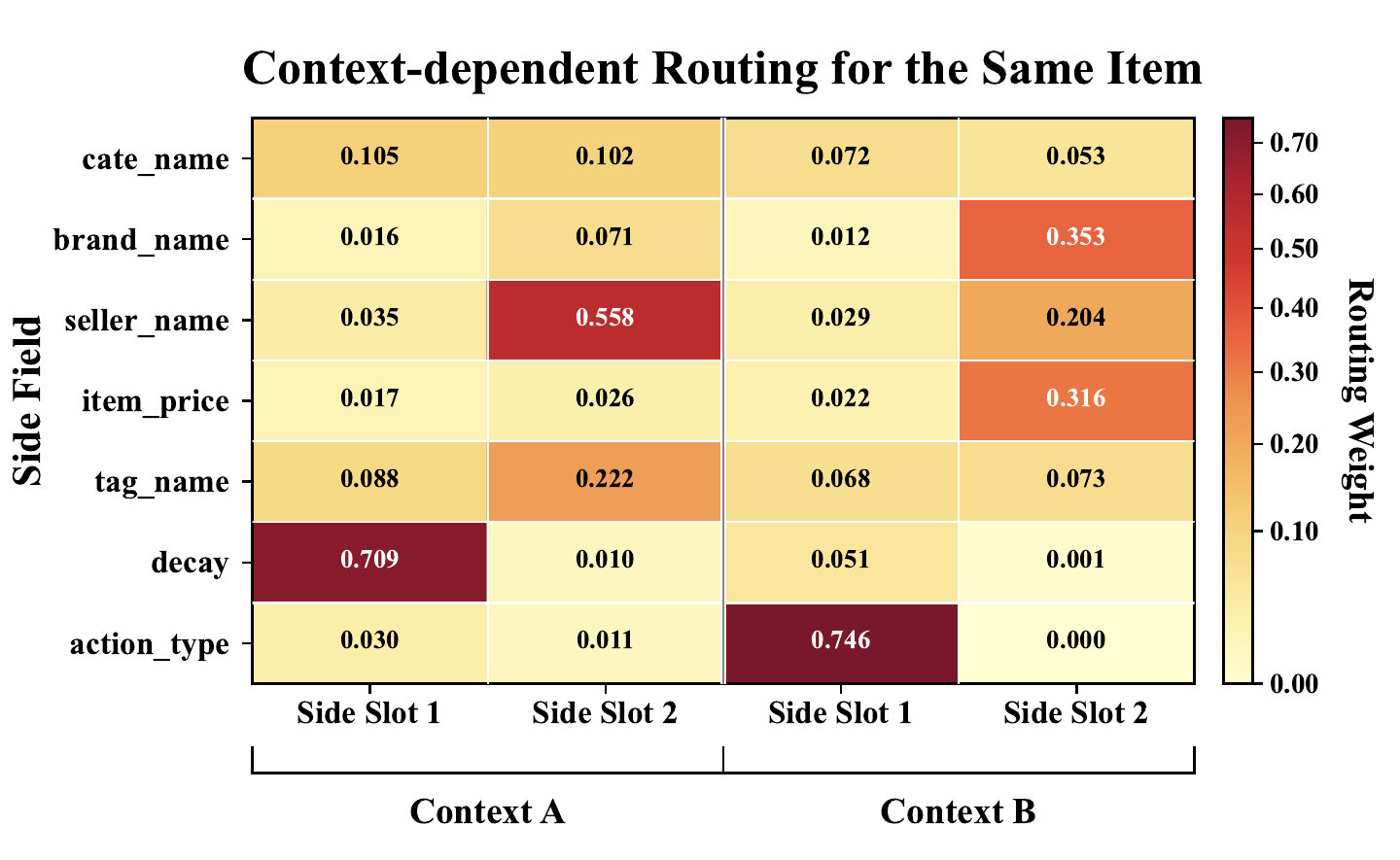}
\caption{Context-dependent side-field routing for the same item. The two contexts correspond to different user histories of the same item, and each column shows the routing distribution of one side slot.}
\label{fig:same_item_context}
\end{figure}
We further analyze whether CAR produces different routing distributions for the same item under different user histories. Specifically, we select an item that appears in two substantially different historical contexts and visualize its side-slot routing weights in each context.

The result is visualized in Figure~\ref{fig:same_item_context}, from which we draw the following observations:
\begin{itemize}[leftmargin=*]
    \item \textbf{The routing distribution of the same item changes across contexts.}
    Although the item identity is fixed, its side-slot routing weights vary substantially under different user histories. For example, Side Slot 1 mainly focuses on \emph{decay} in Context A, while it shifts to \emph{action\_type} in Context B. This shows that CAR does not assign a static side-information representation to an item, but dynamically adjusts the retained fields according to the surrounding behavior sequence.
    \item \textbf{Different slots exhibit distinct field preferences within the same context.}
    Under the same user context, the two side slots also attend to different fields. In Context A, Side Slot 1 concentrates on \emph{decay}, whereas Side Slot 2 assigns higher weights to \emph{seller\_name} and \emph{tag\_name}. A similar pattern appears in Context B, where Side Slot 1 focuses on \emph{action\_type}, while Side Slot 2 emphasizes fields such as \emph{brand\_name}, \emph{item\_price}, and \emph{seller\_name}. This indicates that the side slots learn complementary roles rather than redundant routing patterns.
\end{itemize}

\section{Related Work}
In this section, we review related studies on generative recommendation and heterogeneous behavior modeling.

\textbf{Generative Recommendation.}
Generative recommendation formulates item recommendation as the generation of discrete target item tokens rather than scoring items from a predefined candidate set~\cite{TIGER,P5,EAGER,GenRecLLM}. A typical framework first maps each item into a sequence of semantic IDs (SIDs), represents user histories as token sequences, and then trains an autoregressive model to generate the target item tokens~\cite{TIGER,LETTER,ETEGRec}. This formulation integrates item representation, user modeling, and candidate retrieval within a sequence generation framework.

Existing studies have mainly investigated SID construction and generative modeling. For item tokenization, prior work adopts clustering, vector quantization, residual quantization, learnable tokenization, or end-to-end tokenizer learning~\cite{VQRec,TIGER,LETTER,ETEGRec}. Later studies further incorporate structured attributes, multimodal content, collaborative relations, or behavioral signals into token learning~\cite{EAGER,DAS,MMQ,DiscRec}. In parallel, recent work explores token length, codebook design, decoding efficiency, scaling behavior, and LLM-based generative recommendation~\cite{RPG,GRID,ACERec,GenRecLLM,IntuRec}.

\textbf{Heterogeneous Behavior Modeling.}
User behavior sequences often contain heterogeneous information beyond item IDs or SIDs, including item attributes, categorical features, behavior types, temporal signals, and other contextual features. In sequential recommendation, such information has been incorporated through feature concatenation, additive fusion, attention-based fusion, feature-specific encoders, or multimodal representation learning~\cite{FDSA,S3Rec,NOVA,DIFSR,Grace}. Multi-behavior and time-aware recommendation further model feedback types, behavior dependencies, timestamps, temporal gaps, and recency effects to better characterize user intent~\cite{PBAT,MBSR,TAFAME,DAS}.

When Transformer-based architectures are used for recommendation, representing heterogeneous histories also raises efficiency concerns. Existing methods reduce sequence modeling costs through history truncation, interest aggregation, hierarchical modeling, sparse attention, low-rank attention, or linear attention~\cite{LightSANs,LinRec,Grace}. In generative recommendation, the effective sequence length is further affected by the number of tokens used to represent each historical interaction~\cite{SETRec,RPG,ACERec}. Therefore, how to compactly represent multi-field historical interactions is an important practical issue for generative recommendation.

\section{Conclusion}
In this work, we studied the representation bottleneck of historical interactions in generative recommendation and proposed SPARC, a sequence-aware progressive routing and compression framework. SPARC preserves stable SID-based item identity while dynamically retaining context-dependent side information, enabling each multi-field interaction to be represented as a single compact token under a fixed input budget.

By contextualizing heterogeneous fields before compression, SPARC alleviates the premature information loss caused by static item-wise aggregation. Its progressive design further incorporates field-level and cross-interaction signals without increasing the input length of the generative backbone. Experiments on both industrial and public datasets demonstrate the effectiveness of SPARC and highlight the value of context-aware historical compression for scalable generative recommendation. Future work will explore adaptive token-budget allocation and richer behavioral feature spaces.

\begin{acks}
\end{acks}

\bibliographystyle{ACM-Reference-Format}
\bibliography{9_ref}

\appendix

\section{Dataset}
\label{dataset}
\begin{table}[t]
\caption{Statistical details of the evaluation datasets.}
\label{tab:exp_datastatics}
\resizebox{\columnwidth}{!}{%
\begin{tabular}{ccccc}
\hline
Dataset & \#Users   & \#Items      & \#Interactions & Sparsity \\ \hline
\emph{TaoBao}  & 21 million & 0.27 billion & 26 billion     & 99.99\%  \\
\emph{Beauty}      & 22363        & 12101           & 198502            & 99.93\%        \\
\emph{Toys}      & 19412        & 11924           & 167597             & 99.93\%       \\ \hline
\end{tabular}%
}
\end{table}

\begin{table}[t]
\caption{Component ablation results of SPARC. The best results are highlighted in bold.}
\label{tab:component_ablation}
\resizebox{\columnwidth}{!}{%
\begin{tabular}{ccccc}
\hline
Metric           & SPARC           & \textit{-w/o FCM} & \multicolumn{1}{l}{\textit{-w/o CAR}} & \multicolumn{1}{l}{\textit{-w/o STC}} \\ \hline
$HR^{click}@500$ & \textbf{0.5241} & 0.5227            & 0.5210                                & 0.5208                                \\
$HR^{PV}@500$    & \textbf{0.5648} & 0.5614            & 0.5619                                & 0.5633                                \\ \hline
\end{tabular}%
}
\end{table}

We evaluate the effectiveness of SPARC on both industrial and public datasets. For the industrial dataset, denoted as \emph{TaoBao}, we collect one day of real-world interaction data from TaoBao's production environment on June 16. For the public datasets, we use the Beauty and Toys subsets of the Amazon Review dataset\footnote{https://nijianmo.github.io/amazon/index.html}. Following common practice~\cite{BLOGRec,LETTER,DiscRec}, we apply 5-core filtering to the public datasets and adopt the leave-one-out strategy for data splitting, using the last interaction of each user for testing, the second-to-last interaction for validation, and all preceding interactions for training. For sequence construction, we set the maximum history length to 50; shorter sequences are padded, while longer sequences are truncated to retain the most recent interactions. Detailed dataset statistics are summarized in Table~\ref{tab:exp_datastatics}.

\section{Baselines}
\label{baselines}
The baselines for comparison fall into two categories: conventional and generative recommendation methods.

(a) \emph{Conventional Recommendation Methods}
\begin{itemize}[leftmargin=*]
    \item \textbf{YouTubeDNN}~\cite{YouTubeDNN} represents user preferences by averaging historical item embeddings, producing a compact user vector for efficient candidate retrieval.

    \item \textbf{SASRec}~\cite{SASRec} adopts self-attention within the Transformer architecture to model sequential user behaviors and capture complex item dependencies.

    \item \textbf{BERT4Rec}~\cite{BERT4Rec} adapts the BERT-style masked prediction objective to sequential recommendation, learning contextualized item representations by reconstructing masked items in user sequences.

    \item \textbf{Caser}~\cite{Caser} applies convolutional filters over user interaction sequences to capture both local patterns and broader sequential signals, enabling the modeling of short-term and long-term interests.

    \item \textbf{NextItNet}~\cite{NextItNet} uses dilated convolutional networks to capture long-range sequential dependencies, offering an efficient alternative to recurrent architectures for next-item prediction.

    \item \textbf{CORE}~\cite{CORE} learns session representations through a linear weighted aggregation of item embeddings, aligning session and item representations within the same latent space.
    
\end{itemize}

(b) \emph{Generative Recommendation Methods}
\begin{itemize}[leftmargin=*]
    \item \textbf{HSTU}~\cite{HSTU} formulates retrieval as a sequential transduction task, allowing the model to exploit feature redundancies and improve computational efficiency.

    \item \textbf{TIGER}~\cite{TIGER} is a representative generative retrieval method that employs semantic identifiers and sequence-to-sequence modeling to generate the next item a user may interact with.

    \item \textbf{FORGE}~\cite{FORGE} improves semantic identifier generation by considering collaborative information and ID collision issues, thereby enhancing the effectiveness and utilization of SIDs.

    \item \textbf{RankGR}~\cite{RankGR} enhances generative retrieval with listwise direct preference optimization to model hierarchical user preferences and employs a lightweight scoring module to refine top-ranked candidates through interactions with user behavior sequences.
\end{itemize}

\section{Component Ablation}

To further investigate the contribution of each component in SPARC, we conduct an component ablation study by individually removing FCM, CAR, and STC. The results are reported in Table~\ref{tab:component_ablation}. We evaluate all variants on both $HR^{click}@500$ and $HR^{PV}@500$. From the results, we can observe
that removing any individual component leads to a certain degree of performance degradation, demonstrating the effectiveness and necessity of each component in the overall design of SPARC.

\end{document}